\documentclass{aastex61}

\newcommand\gsim{\,\lower3pt\hbox{$\sim$}\llap{\raise2pt\hbox{$>$}}\,}
\newcommand\lsim{\,\lower3pt\hbox{$\sim$}\llap{\raise2pt\hbox{$<$}}\,}

\usepackage{graphicx}
\usepackage{epsfig}
\usepackage{natbib}

\submitjournal{Astrophysical Journal Letters}

\shortauthors{LUGAZ ET AL.}

\shorttitle{MULTI-SPACECRAFT CME MEASUREMENTS AT SCALES OF 0.01 AU}

\begin{document}

%
%

\title{On the Spatial Coherence of Magnetic Ejecta: Measurements of Coronal Mass Ejections by Multiple Spacecraft Longitudinally Separated by 0.01 AU}

\author[0000-0002-1890-6156]{No{\'e}\ Lugaz}
\affiliation{Space Science Center, Institute for the Study of Earth, Oceans, and Space, University of New Hampshire, Durham, NH, USA}
\affiliation{Department of Physics, University of New Hampshire, Durham, NH, USA}
\author{Charles~J. Farrugia}
\affiliation{Space Science Center, Institute for the Study of Earth, Oceans, and Space, University of New Hampshire, Durham, NH, USA}
\affiliation{Department of Physics, University of New Hampshire, Durham, NH, USA}
\author{Reka~M. Winslow}
\affiliation{Space Science Center, Institute for the Study of Earth, Oceans, and Space, University of New Hampshire, Durham, NH, USA}
\author{Nada~Al-Haddad}
\affiliation{Institute for Astrophysics and Computational Sciences, Catholic University of America, Washington, DC, USA}
\author{Antoinette~B.~Galvin}
\affiliation{Space Science Center, Institute for the Study of Earth, Oceans, and Space, University of New Hampshire, Durham, NH, USA}
\affiliation{Department of Physics, University of New Hampshire, Durham, NH, USA}
\author{Teresa~Nieves-Chinchilla}
\affiliation{Institute for Astrophysics and Computational Sciences, Catholic University of America, Washington, DC, USA}
\affiliation{NASA/Goddard Space Flight Center, Greenbelt, MD, USA}
\author{Christina~O.~Lee}
\affiliation{Space Sciences Laboratory, University of California, Berkeley, CA, USA}
\author{Miho Janvier}
\affiliation{Institut d'Astrophysique Spatiale, CNRS, Univ.\ Paris-Sud, Universit{\'e} Paris-Saclay, France}

%
%

\begin{abstract}
Measurements of coronal mass ejections (CMEs) by multiple spacecraft at small radial separations but larger longitudinal separations is one of the ways to learn about the three-dimensional structure of CMEs. Here, we take advantage of the orbit of the {\it Wind} spacecraft that ventured to distances of up to 0.012 astronomical units (au) from the Sun-Earth line during the years 2000 to 2002. Combined with measurements from ACE, which is in a tight halo orbit around L1, the multipoint measurements allow us to investigate how the magnetic field inside magnetic ejecta (MEs) changes on scales of 0.005--0.012~au. We identify 21 CMEs measured by these two spacecraft for longitudinal separations of 0.007 au or more.  We find that the time-shifted correlation between 30-minute averages of the non-radial magnetic field components measured at the two spacecraft is systematically above 0.97 when the separation is 0.008 au or less, but is on average 0.89 for greater separations. Overall, these newly analyzed measurements, combined with 14 additional ones when the spacecraft separation is smaller, point towards a scale length of longitudinal magnetic coherence inside MEs of 0.25 -- 0.35~au for the magnitude of the magnetic field but 0.06 -- 0.12~au for the magnetic field components. This finding raises questions about the very nature of MEs. It also highlights the need for additional ``mesoscale'' multi-point measurements of CMEs with longitudinal separations of 0.01 -- 0.2 au. 
\end{abstract}

\section{INTRODUCTION} \label{intro}

Historically, multi-spacecraft measurements of the same coronal mass ejection (CME) have been one of the main ways to understand the structure of these eruptions. In fact, the seminal article by \citet{Burlaga:1981} that defined a magnetic cloud (MC) as a subset of CMEs with well defined properties is based on measurements of the same CME made by five different spacecraft, three of which 
were at about the same radial distance from the Sun. The evolution of CME properties with radial distances is primarily known from statistical studies of different CMEs measured at different distances, under the assumption that the average CME behavior can be determined from studying a large enough sample of CMEs \citep[]{Liu:2005,Leitner:2007,Winslow:2015}. This has been complemented by a few case studies of the same CMEs measured by different spacecraft, primarily with MESSENGER  and spacecraft near 1 AU \citep[]{Winslow:2016,Winslow:2018}, MESSENGER, Venus Express (VEX), Pioneer Venus Orbiter (PVO) and spacecraft near 1 AU \citep[]{Jian:2008, Good:2015,Good:2018, Wang:2018}, NEAR and spacecraft near 1 AU \citep[]{Mulligan:1999, Mulligan:2001}. These measurements have so far confirmed the results from statistical and theoretical studies, and focused on the dependence on radial distance or time of the CME radial size, speed, magnetic field strength and magnetic field orientation.  \cite{Bothmer:1998} and \citet{Good:2016} also learned about the CME longitudinal size by studying how multiple spacecraft measuring the same CME depends on the spacecraft separation. 

Because CME properties evolve significantly as they propagate \citep[e.g., see][]{Manchester:2017}, it is most appropriate to learn about their three-dimensional properties from measurements by multiple spacecraft at nearly the same heliocentric distance. \citet{Janvier:2013}, \citet{Demoulin:2013} and \citet{Janvier:2015} performed a series of investigation of the shape of MCs and shocks based on the properties of the nearly 150 MCs and 300 shocks measured near L1 by the {\it Advanced Composition Experiment} (ACE) and {\it Wind} since 1995. One of the goals of the STEREO mission \citep[]{Kaiser:2008} is to provide multi-point measurements of transient phenomena from different vantage points near 1~AU. However, the mission was launched during the deep solar minimum of solar cycle 23 (SC23) and there were very few multi-spacecraft measurements of CMEs near 1~au that involved  STEREO \citep[]{Kilpua:2009b, Liu:2008b,Farrugia:2011}. STEREO has nonetheless revolutionized our understanding of another type of transient event through widespread measurements of solar energetic particles (SEPs) by spacecraft separated by up to 180$^\circ$ in longitude \citep[]{Dresing:2012,Lario:2013}. 
A very comprehensive review of multi-spacecraft measurements before and during the first few years of STEREO can be found in \citet{Kilpua:2011} and, thus, we will not repeat these here. The closest angular separation between two spacecraft measuring the same CME was 3$^\circ$, corresponding to about 0.05~au. One of the events studied by \citet{Mulligan:1999} was for a radial separation of 0.21~au and an angular separation of 1$^\circ$. We note that 0.0175~au $=$ 410~$\mathrm{R_E}$ $=$ 3.76~$\mathrm{R_\odot}$ corresponds to 1$^\circ$ in heliolongitude separation for two spacecraft near 1~au.

It is interesting to compare these separations to the expected size of a ME at 1~au. We take the ME radial width as 0.21~au following \citet{Lepping:2005}. The cross-section can be assumed to be circular or elliptical based on a study of the distribution of the impact parameters by \citet{Demoulin:2013}. For a circular cross-section, the ME half-angle at 1~au is about 6$^\circ$. The third dimension, the extent of the CME axis, is typically estimated to be a half-angle of 30-40$^\circ$ \citep[]{Janvier:2015}.

From September 2000 to July 2002, {\it Wind} performed distant prograde orbits reaching up to 320~$\mathrm{R_E}$ from ACE in the Geocentric Solar Ecliptic (GSE) $y$-direction while typically remaining within 30~$\mathrm{R_E}$ from ACE in the GSE $z$-direction. The nearly two years of measurements near the maximum of SC23 provide a unique opportunity to study how the properties inside MEs differ at longitudinal separations between 0.006 and 0.012~au. 
\citet{Koval:2010} took advantage of this longitudinal separation to study the radius of curvature of interplanetary shocks, while other studies focused on the correlation scales in the solar wind and turbulence \citep[]{Wicks:2009, King:2005, Ogilvie:2007}. No similar study has been performed for CMEs. There has been a couple case studies on correlation lengths inside CMEs \citep[]{Matsui:2002,Farrugia:2005} that focused primarily on the coherence of the measurements with increasing separation in the radial direction. 

Here, we perform a dedicated study of the magnetic field correlation between ACE and {\it Wind} from September 2000 to May 2002, for non-radial separations of the order of 0.01~au. The rest of the article is organized as follows. In Section~\ref{sec:data}, we discuss our data selection and methodology. In Section~\ref{sec:examples}, we discuss one example before presenting the full results in Section~\ref{sec:corr}. We discuss our results, especially in terms of length scales in MEs and the need for future missions in Section~\ref{sec:discussions}. We conclude in Section~\ref{sec:conclusion}.

\section{Data Selection and Methodology}\label{sec:data}

We use the {\it Magnetic Field Investigation} (MFI) onboard {\it Wind} and ACE \citep[]{Lepping:1995,Smith:1998}. We focus on clear MEs measured as the ACE-{\it Wind} separation in the $y$-direction was at least 165~$\mathrm{R_E}$ (7\,$\times 10^{-3}$~au). We start from the list of all MCs identified by \citet{Lepping:1990} or listed as such in \citet{Richardson:2010} as well as a few clear MC-like events ({\it i.\,e}.\ listed as ``1'' in \citet{Richardson:2010}), for example the two MEs part of the 2001 March 31-April 1 multiple-MC event \citep[]{Wang:2003,Farrugia:2004}. Overall, we identify 22 such MEs, one of which was found to be a very poor event and removed from the analysis (2002 May 23). In addition to these 21 MEs, we select 14 additional clear MEs to get multipoint measurements for spacecraft with longitudinal separations between 32 and 132~$\mathrm{R_E}$. We refer to the ejecta part of these CMEs as MEs, whether they are MCs or MC-like. 

For these 35 CMEs, we apply the following analysis procedure: 1) we use 3-second {\it Wind} data and 16-second ACE data in the Geocentric Solar Ecliptic (GSE) coordinate system. We tested the importance of data of higher time resolution by using 1-second ACE data, and found no significant difference. 2) We bin the data into 30-second averages (using the resampling function of python pandas). 3) We select the ME boundaries using the information provided by the ICME database of \citet{Richardson:2010}, that of \citet{Jian:2006b} and \citet{Nieves:2018} and the MC lists of \citet{Lepping:1990} and \citet{Huttunen:2005}. Typically, we choose boundaries larger than that of the MCs, comparable to that of the magnetic obstacle of the list of \citet{Nieves:2018} and less than that of the ICME Plasma/Field boundaries of \citet{Richardson:2010}. These boundaries are used for the rest of the analysis. 4) We calculate the lag between the {\it Wind} and ACE magnetic field data inside the ME that maximizes the Pearson correlations between these measurements. We maximize the correlation and calculate different lags for the magnetic field strength, and each of the three components, separately. 5) We rebin the data into 30-minute averages. 30-minute is the time-resolution typically used for ME fitting and reconstruction \citep[]{Lepping:1990}. For typical CME speeds, 30-minute data corresponds to a radial length between 0.5 and 1.0 $\times 10^{-2}$~au. As such, we focus on the large-scale structure of the ME or sub-structures that are at least as long in the radial as in the longitudinal directions. 

\section{Specific Example: 2002 May 19 ME}\label{sec:examples}

We highlight the 2002 May 19 CME, for which the ACE-{\it Wind} separation was (9.58, 11.8, 1.4 )~$\times 10^{-3}$~au in the three GSE components. As will be shown later, this can be considered a typical example of the behavior of the magnetic field inside a ME at longitudinal separations of about 0.01~au. 

\begin{figure*}[htb]
\centering
{\includegraphics*[width=10cm]{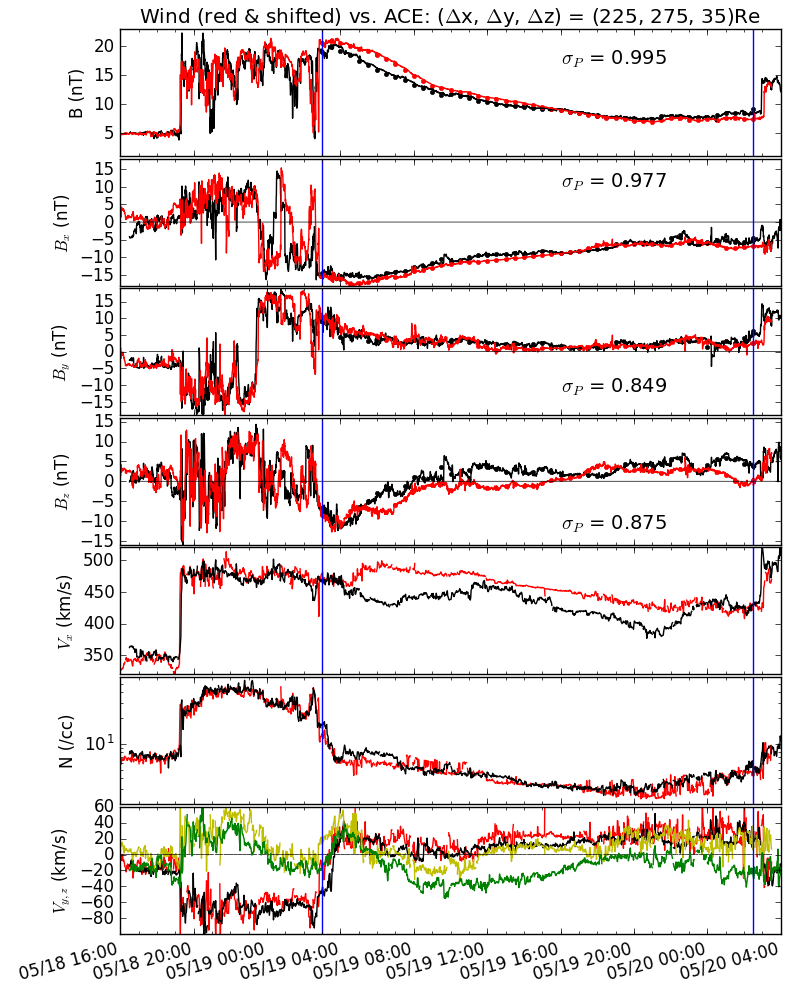}}
\caption{Comparison of ACE (black) and {\it Wind} (red, time-shifted) measurements for the 2002 May 19 CME. The panels show from top to bottom, the total magnetic field strength, the $x$, $y$  and $z$ components of the magnetic field vector in GSE coordinates, the proton radial velocity, density (log scale) and the $y$ (black/red) and $z$ (green/yellow) components of the velocity. The blue vertical lines delimit the ME boundaries, where the correlation coefficients for the binned data (dots) are calculated.}
\label{example}
\end{figure*}

The 2002 May 19 event, shown in Figure~\ref{example}, is a relatively clear MC of moderate speed (450~km\,s$^{-1}$). ACE observes a fast forward shock at 19:19 UT on May 18. Wind measures the same shock at 19:46 UT. The shock speed and magnetosonic Mach number are very similar at ACE and {\it Wind}: speed of 530~$\pm~10$~km\,s$^{-1}$ and  Mach number of 4~$\pm~0.2$.
A MC is listed by \citet{Lepping:1990} from 3.9 to 23.4~UT on May 19 and from 4 to 22~UT in the list of \citet{Huttunen:2005}. The magnetic obstacle/ejecta is listed from 02:40 on May 19 to 02:57 UT on May 20 in \citet{Jian:2006b} based on ACE data and from 03:30 on May 19 to 03:34 UT on May 20 in \citet{Nieves:2018} based on {\it Wind} data. The end of the ME is determined by the arrival of another fast-forward shock at 02:57/03:36 UT on May 20 (ACE/{\it Wind} timing). For the correlation, we use data from 3UT on May 19 to 2:30UT on May 20. 

The magnetic field correlation coefficients between ACE and {\it Wind} are 0.995, 0.977, 0.849 and 0.875 for the magnitude and the $x$, $y$ and $z$ components of the magnetic field, respectively. Throughout the ME, there are differences in both plasma and magnetic field measurements. There are some differences in the magnetic field strength from 4 to 10 UT on May 19 of up to 2.5 nT ($\sim 10-15\%$). The largest deviations are for $B_z$ with variations of up to 5~nT. In particular, starting at 09:40 UT on May 19 for 6 hours, {\it Wind} measures small southward magnetic fields, whereas ACE measures larger northward magnetic fields.  Note that the accuracy of ACE and {\it Wind} magnetometers is of the order of $\pm~0.1$~nT \citep[]{Lepping:1995,Smith:1998}, and the differences measured here are at least one order of magnitude larger.

This same time period is also characterized by differences in the $x$ and $z$ components of the velocity, with {\it Wind} measuring faster flows with a small northward components and ACE slower flows with a stronger ($-30$~km\,s$^{-1}$) southward component. The difference in the velocity measured by {\it Wind} and ACE reaches 60~km\,s$^{-1}$. 
After the second shock arrival, the velocity matches much better between the two spacecraft, confirming that the difference is not an instrumental effect. 
This MC is found to be fitted fairly by \citet {Lepping:1990} (quality flag of 1); it has a low inclination ($i = 8^\circ$) and is crossed close to its ``nose'' ($\lambda = 14^\circ$), but the impact parameter is high (0.95), meaning it is crossed far from its axis. With this inclination, ACE and {\it Wind} are aligned parallel to the MC axis and are expected to observe very similar signatures. The MC radial width is estimated from the fitting to be 0.42~au or about 28 times larger than the ACE-{\it Wind} separation. Even at a separation less than 5\% of the ME width, {\it Wind} and ACE measure different magnetic fields inside the ME. 

\section{Results for all 35 CMEs}\label{sec:corr}

\begin{figure*}[htb]
\centering
{\includegraphics*[width=8.7cm]{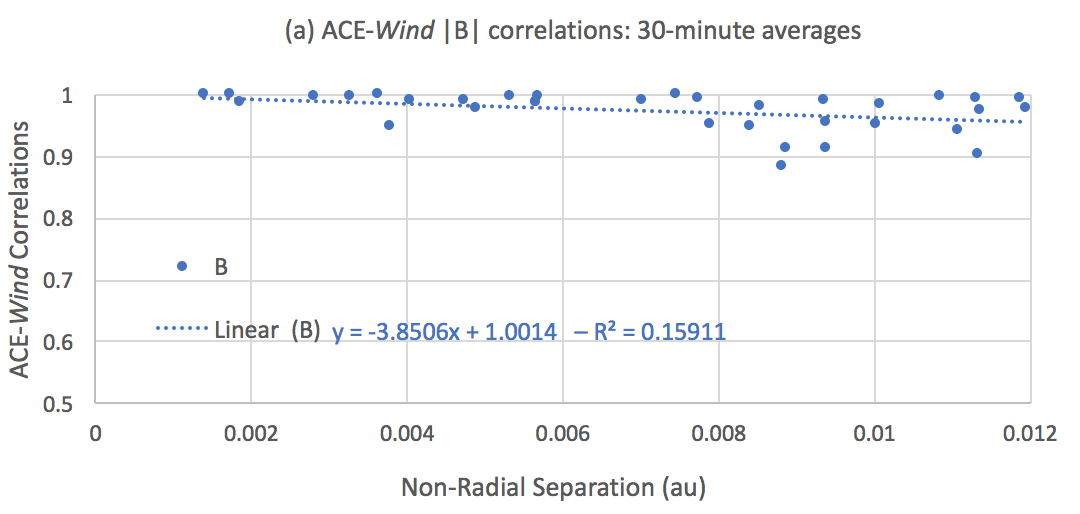}}
{\includegraphics*[width=8.7cm]{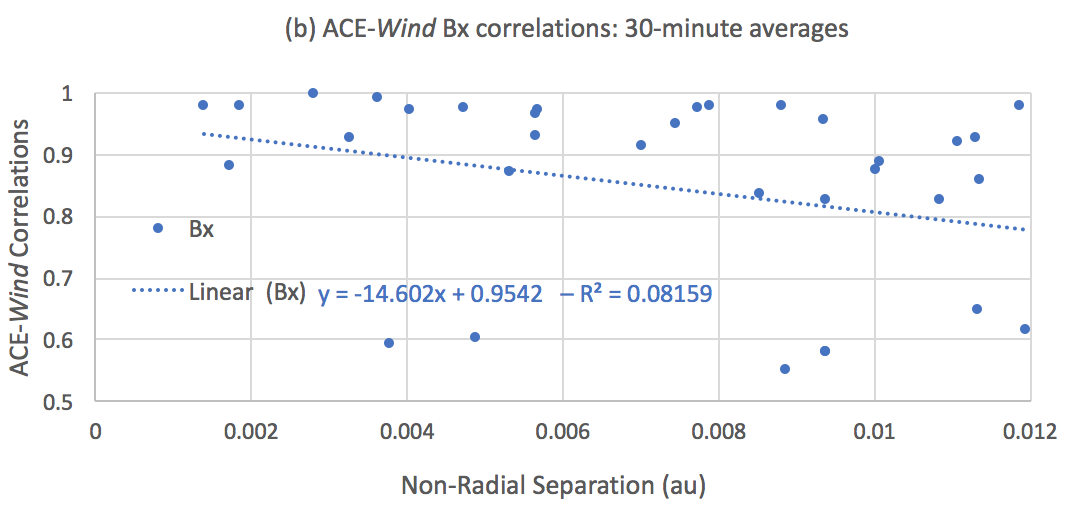}}\\
\vspace{0.5cm}
{\includegraphics*[width=8.7cm]{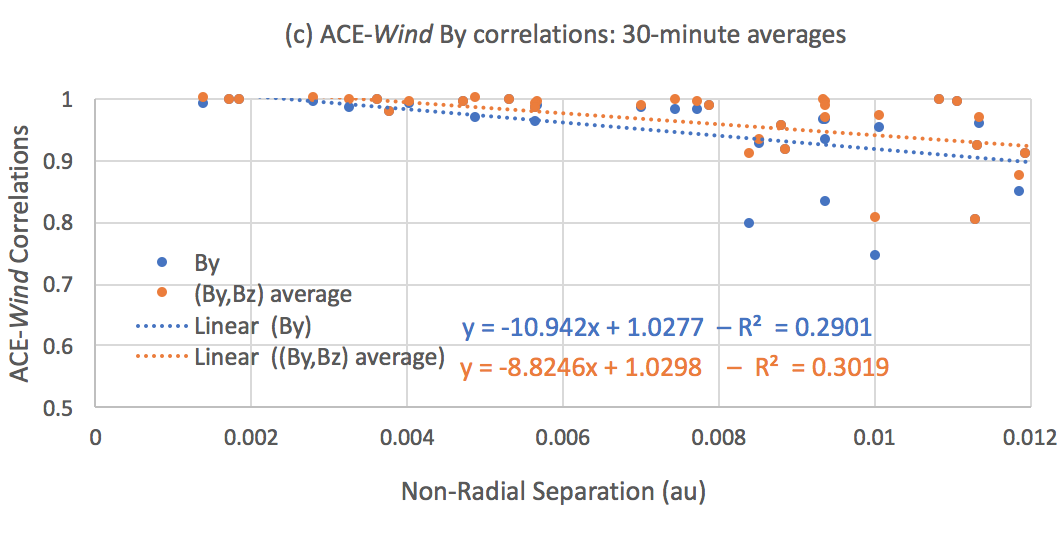}}
{\includegraphics*[width=8.7cm]{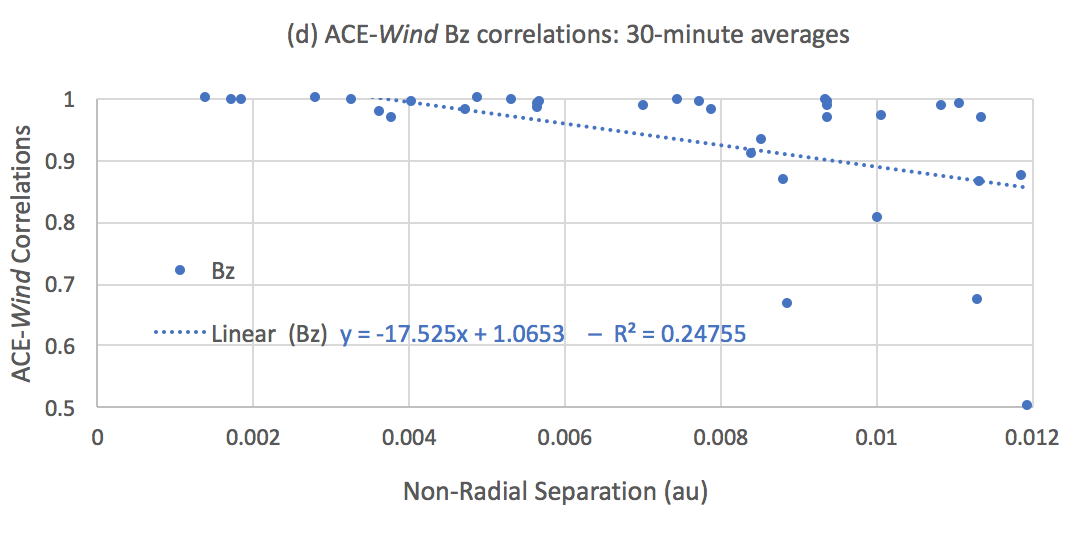}}
\caption{Correlation coefficients of the measurements at ACE and {\it Wind} {\it vs.}\ non-radial separation for the magnetic field magnitude (panel a) and $x$, $y$ and $z$ components (panels b, c and d). Panel (c) also includes the average (plotted in orange) of the correlation coefficients for $B_y$ and $B_z$. In each panel, the linear fit to the data is plotted and the linear best-fit equation and $R^2$ values are included.}
\label{fig:corr}
\end{figure*}

Figure~\ref{fig:corr} shows the Pearson correlation coefficients of the time-lagged {\it Wind} and ACE magnetic field measurements for the 35 studied CMEs as a function of the absolute value of the non-radial ($y$ and $z$ in GSE) separation between ACE and {\it Wind}. Supplementary Table~1 lists all 35 CMEs, the boundaries used, the correlation coefficients and the MC orientation, when available. We note that the correlation coefficient for the $B_x$ component of the magnetic field is often constrained by the lack of variation in that component inside the ME which makes the correlation coefficient ill-defined. We plot it nonetheless for completeness. For separations of 0.003~au or smaller, the correlations are above 0.98, highlighting that ACE and {\it Wind} observe the same fields when closely aligned.  All correlation coefficients show a relatively clear decreasing trend with increasing separation. Separating the measurements into two groups, for separations of less or more than 0.008~au, the correlation coefficients of these two groups are statistically different at least at the 95\% level, with P-values between 0.013 for $B_x$ and 0.0001 for $B_y$. 

We test different correlations between the MC parameters or the spacecraft separation and the correlation coefficients of the ACE and {\it Wind} magnetic field measurements. 
The two clearest relations are between the correlation of the $B_y$ and $B_z$ components and the non-radial separation of the spacecraft with a correlation of $-0.539$ and $-0.498$. 
There is no correlation that reaches 0.3 or above between $B_y$ or $B_z$ and the MC parameters, including the axis inclination and impact parameter. This may be due to the small sample size, or low and high-inclination MEs have similar changes with non-radial distances. 
The best correlations for the $B_x$ component are found for the MC inclination (0.456), the $x$ (0.327) and total (0.328) spacecraft separations. For the total magnetic field, the best correlations are for the total (0.456) and non-radial (0.401) separations. Taking the average of the $B_y$ and $B_z$ correlations, the correlation of the average non-radial magnetic field with the non-radial separation reaches $-0.575$. 

\section{Discussions}\label{sec:discussions}

\subsection{Consequences for Scale Length of Magnetic Ejecta}

We can use the information from the correlation in the magnetic field measurements between two separated spacecraft to obtain a physics-based estimate of the scale length inside MEs. This topic has been recently highlighted by the work of \citet{Owens:2017} that postulated that CMEs are not coherent structures. Our work proposes an alternate theory, as highlighted below.

We perform a linear fit of the {\it Wind}-ACE magnetic correlation coefficients with the non-radial separation as shown in Figure~\ref{fig:corr}. It is then extrapolated to determine the distance at which zero correlation is expected. We refer to this distance thereafter as the length scale of the magnetic field (component). We also test the importance of the averaging window by calculating the scale length for averages of 30 seconds as well as 15 and 60 minutes. We find that the scale lengths for $B_y$, $B_z$, their average correlation and the maximum of their correlation are about 3, 4.5, 3.5 and 2.5 times less than the scale length of $B$, for every averaging duration.  
The scale length is found to vary from 0.26 $\pm 0.1$~au for the magnetic field magnitude to 0.094 $\pm 0.02$ and 0.061 $\pm 0.002$~au for the $B_y$ and $B_z$ components, respectively. Using the average (resp.\ maximum) of the $B_y$ and $B_z$ correlations, the scale length for the non-radial magnetic field is found to be 0.0735 $\pm 0.007$~au (resp.\ 0.117 $\pm 0.015$~au). Uncertainties are based on the variation found using averages of 15, 30 and 60 minutes.

The typical radial width of a ME at 1~au is 0.21~au based on fittings; our result for the correlation between ACE and {\it Wind} magnetic field magnitude points towards a similar size in the non-radial direction, on average. The fact that there is no influence of the ME orientation, including inclination, might be due to the relatively small number of events. However, it is clear that the magnetic field $B_y$ and $B_z$ components behave differently than the magnitude, with a much faster decrease in correlation with increasing separations. Overall, this points towards MEs having two significant scales near 1~au: one related to the magnetic field components which is 0.07--0.12~au (4--7$^\circ$) and one for the total magnetic field which is 0.25--0.35~au (14--20$^\circ$).

\subsection{Multi-spacecraft Measurements of CMEs: A Gap in Our Knowledge of CMEs}

\begin{figure*}[htb]
\centering
{\includegraphics*[width=9cm]{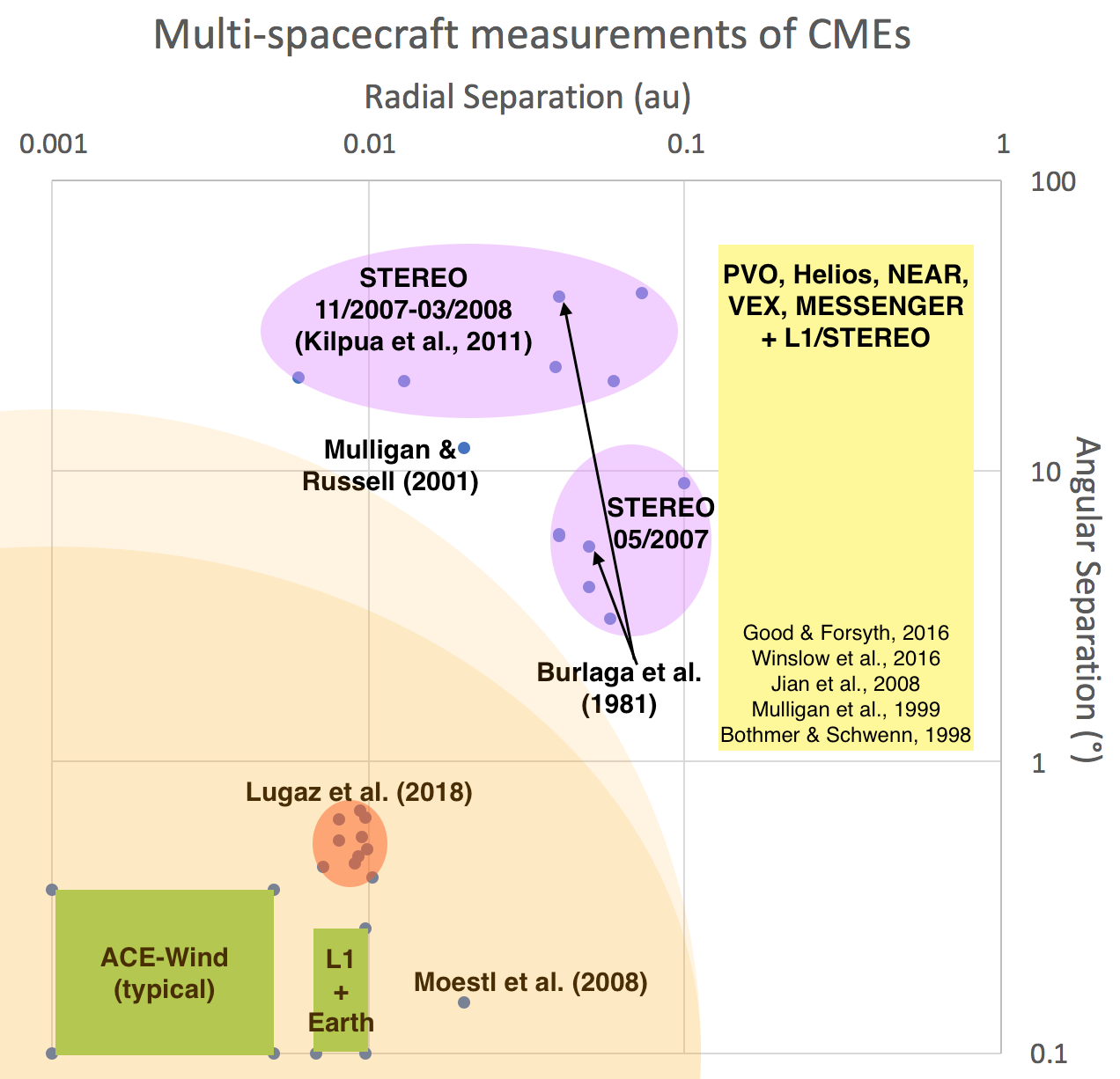}}
\caption{Past multi-spacecraft measurements of CMEs including the present work. Overlaid in orange is the expected cross-section of a ME (light: 0.11 AU radius and 16$^\circ$, darker: 0.11~au radius and 5.5$^\circ$). The critical region of 1-12$^\circ$ separation has not been sampled except for large radial separations.}
\label{fig:before}
\end{figure*}

In order to illustrate regions of unexplored parameter space, it can be enlightening to plot past multi-spacecraft CME measurements on a two-dimensional grid, in terms of radial and longitudinal separations, as shown in Figure~\ref{fig:before}. We do not focus on radial separations beyond 1~AU as made possible by Voyager and Ulysses. The right-hand side of this Figure corresponds to large radial separations and a variety of longitudinal separations. It is populated by near-conjunction studies, primarily from planetary missions. The bottom left corner is populated by near-simultaneous measurements by ACE and {\it Wind}. Because ACE is on a tighter halo orbit around L1 than {\it Wind} is, the radial separation varies between 0 and 50~$R_\mathrm{E}$ and the longitudinal separation varies between 0 and 150~$R_\mathrm{E}$, except in 1999--2003. The rest of the plot is populated by the few STEREO measurements, the measurements of \citet{Burlaga:1981}, \citet{Mulligan:1999} and the study of \citet{Moestl:2008} with {\it Wind} in Earth's distant magnetotail.  The region of the parameter space left unexplored (radial separations of 0.005 to 0.05~au and longitudinal separations of 1 to 12$^\circ$) corresponds to the expected size of the cross-section of a ME (shown in orange). The current study highlights that this mesoscale region may hold the key to determine the magnetic structure of CMEs.

\subsection{Recommendations}
The gap identified in the previous section needs to be filled if we are to make progress in our understanding of CMEs, the main drivers of intense space weather. We propose three specific ways to fill this gap. All three suggestions are likely to be necessary for us to understand the longitudinal structure of CMEs.

1- All efforts should be made to prolong the STEREO mission at least to the end of 2024. From 2023 March 20 to 2024 May 1, the separation between Earth and STEREO-A will be less than 12$^\circ$, allowing once again for multi-spacecraft measurements of CMEs. Measurements at separations under 6$^\circ$ are especially important based on the current study. This will be possible for only 4.5 months from 2023 June 15 to 2023 November 10. If STEREO-B can be recovered, it will provide 1.5 additional months at these mesoscale separations (from May 5 onwards). 

2- With the presence of ACE and DSCOVR in a tight L1 orbit, it would be highly beneficial for space physics to have the {\it Wind} spacecraft going back into a prograde orbit, fuel-permitting. This would enable additional multi-spacecraft measurements of shocks and CMEs at small to moderate separations ($\sim$ 0.01~au). Ideally, a dedicated campaign with {\it Wind} in a prograde orbit, ACE/DSCOVR at L1 and STEREO-A would provide 3-point measurements of CMEs around 2023 and could provide an exceptional ``grand finale'' for {\it Wind}.

3- In light of the short time span when STEREO will be close enough to L1 to make multi-spacecraft measurements, a dedicated mission to investigate heliospheric mesoscales is required. Such a mission could be relatively simple, with only magnetic field, plasma and SEP measurements required. A H{\'e}non distant retrograde orbit \citep[]{Henon:1997}, at closer separations from L1  than the one discussed by \citet{StCyr:2000}, would fit the scientific requirements for such a mission. In addition to learning about CMEs, it could give new and yet unavailable information about shocks, corotating streams and the local acceleration of particles.   

\section{Conclusions}\label{sec:conclusion}

We identify 21 CMEs with clear ME measured by ACE and {\it Wind} when the non-radial separation of these spacecraft was at least 0.008~au. We find that the correlation between 30-minute averages of the non-radial magnetic field components measured by ACE and {\it Wind} is on average 0.91 at these separations with a trend indicating that zero-correlation would be reached at separations of 0.08~au. The quick decrease in correlation in the $B_y$ and $B_z$ components for separations that are much smaller than the expected ME size is a finding that points towards a necessary reconsideration of the nature of MEs. It could be due to MEs being composed of a small coherent core of about 0.1~au surrounded by a larger region as discussed by \citet{Savani:2013} or related to the technique of \citet{Mulligan:2013}. It could also be associated with a truly 3-D structure without axial invariance \citep[]{Jacobs:2009,AlHaddad:2011}. These findings point towards a gap in observations and knowledge about CMEs that should be filled with new missions.

\begin{acknowledgments}

The research for this manuscript was supported by the following grants: NSF AGS-1435785, AGS-1433086 and AGS-1433213 and NASA NNX15AB87G,  NNX15AU01G and NNX16AO04G. R.M.W. acknowledges support from NASA grant NNX15AW31G and NSF grant AGS1622352.

\end{acknowledgments}

\bibliographystyle{apj}

\end{document}